\documentclass[a4paper,11pt]{article}
\pdfoutput=1 
\usepackage{amssymb}
\usepackage{jcappub} 

\usepackage[T1]{fontenc} 
\hypersetup{
     colorlinks   = true,
     citecolor    = blue,
     urlcolor     = blue,
     linkcolor    = blue
}
\usepackage[dvipsnames]{xcolor}
\usepackage{verbatim}
\usepackage{supertabular}
\usepackage{graphicx}
\usepackage{subcaption}
\usepackage{longtable}

\definecolor{aquamarine}{rgb}{0.2,0.7,0.6}

\title{\boldmath Seeking the nearest neutron stars using a new local electron density map}

\author[a,b,c]{Joseph Bramante,}
\author[c]{Katherine Mack,}
\author[e]{Nirmal Raj,}
\author[f,g]{Lijing Shao,}
\author[a,b,1]{Narayani Tyagi\note{Corresponding author}}

\affiliation[a]{Department of Physics, Engineering Physics, and Astronomy, Queen’s University, Kingston, Ontario, K7N 3N6, Canada}
\affiliation[b]{The Arthur B. McDonald Canadian Astroparticle Physics Research Institute, Kingston, Ontario, K7L 3N6, Canada}
\affiliation[c]{Perimeter Institute for Theoretical Physics, Waterloo, Ontario, N2L 2Y5, Canada}
\affiliation[e]{Centre for High Energy Physics, Indian Institute of Science, C.V. Raman Avenue, Bengaluru 560012, India}
\affiliation[f]{Kavli Institute for Astronomy and Astrophysics, Peking University, Beijing 100871, China}
\affiliation[g]{National Astronomical Observatories, Chinese Academy of Sciences, Beijing 100012, China}

\emailAdd{joseph.bramante@queensu.ca}
\emailAdd{kmack@perimeterinstitute.ca}
\emailAdd{nraj@iisc.ac.in}
\emailAdd{lshao@pku.edu.cn}
\emailAdd{narayani.tyagi@queensu.ca}

\abstract{Neutron stars provide a compelling testing ground for gravity, nuclear dynamics, and physics beyond the Standard Model, and so it will be useful to locate the neutron stars nearest to Earth.
To that end, we revisit pulsar distance estimates extracted from the dispersion measure of pulsar radio waves scattering on electrons. 
In particular, we create a new electron density map for the local kiloparsec by fitting to {\em parallax} measurements of the nearest pulsars, which complements existing maps that are fit on the Galactic scale.
This ``near-Earth'' electron density map implies that pulsars previously estimated to be 100$-$200 pc away may be as close as tens of parsecs away, which motivates a parallax-based measurement campaign to follow-up on these very-near candidate pulsars. 
Such nearby neutron stars would be valuable laboratories for testing fundamental physics phenomena, including several late-stage neutron star heating mechanisms, using current and forthcoming telescopes. 
We illustrate this by estimating the sensitivities of the upcoming Extremely Large Telescope and Thirty Meter Telescope to neutron stars heated by dark matter capture.}

\begin{document} 

\maketitle
\flushbottom

\section{Introduction}

How near is the nearest neutron star to Earth? 
Answering this question accurately may have far-reaching implications for fundamental physics and astrophysics, since neutron stars (NSs) constitute some of their most sensitive laboratories~\cite{NSfundamental:Nattila:2022evn,BramanteRajCompactDark:2023djs,Shao:2022koz}. 
For example, precise measurements of the masses and radii of nearby NSs would be essential to constrain the state equation of high-density matter~\cite{Lattimer:2000nx}, and their velocities would help pinpoint their kinematic age~\cite{kinematicage2010}, all of which would sharpen our understanding of passive cooling of NSs~\cite{cooling:Yakovlev:2004iq,coolingminimal:Page:2004fy,coolingcatalogue:Potekhin:2020ttj}.
Of particular interest to us in this regard are the late-stage reheating mechanisms of NSs, such as numerous heating mechanisms involving a hidden sector of particles~\cite{BramanteRajCompactDark:2023djs}, including dark matter and other proposed astrophysical effects~\cite{NsvIR:otherinternalheatings:Reisenegger}.
For recent reviews of hidden sector and astrophysical mechanisms, see $e.g.$ Ref.~\cite{BramanteRajCompactDark:2023djs} and the Appendix of Ref.~\cite{Raj:2024kjq}.

Only about $5-10\%$ of NSs are observed as radio pulsars, primarily due to their beam orientation, brightness, and distance limitations.
Including x-ray and gamma-ray observations, the observable fraction might be closer to $10\%$~\cite{Lorimer:2010pa}.
The closest known pulsar is estimated to be 110$-$130~pc away~\cite{closestNSparallaxHSTWalter:2010ht,closestNS2022arXiv221004685Z}, whereas from the number density of NSs in the solar vicinity, $n_\odot \simeq (1-5) \times10^{-4} {\rm pc}^{-3}$ (assuming $10^9$ NSs in the Galaxy)~\cite{1993ApJ...403..690B,NSdistribs:Sartore2010}, we obtain a theoretical distance to the nearest NS of only $(3/4\pi n_\odot)^{1/3} \simeq 10$~pc.
This calls for a close scrutiny of the region around us. 
Of course, the spatial pulsar distribution in our galaxy has been a topic of interest for a few decades now. Thanks to the current generation of large-scale pulsar surveys~\cite{Manchester_2005,2018ApJ...859...93L,GBNCC2014ApJ...791...67S,2021ApJ...922...35A,2020ApJ...892...76M,2019ApJ...875...19A,2018ApJ...857..131K,1997jena.confE.236L}, we now have large samples of both regular ($\mathcal{O}(1)$~s period) and millisecond pulsars. 
 Along with the ATNF pulsar catalogue~\cite{Manchester_2005}, which is considered the standard, other databases such as the EPN database of pulsar profiles~\cite{1997jena.confE.236L} and the Green Bank North Celestial Cap (GBNCC) pulsar survey~\cite{2018ApJ...859...93L,GBNCC2014ApJ...791...67S,2021ApJ...922...35A,2020ApJ...892...76M,2019ApJ...875...19A,2018ApJ...857..131K}, provide detailed accounts of discovered pulsars.

Pulsar distance estimates in these catalogues are often based on radio pulsar dispersion measures (DMs), which in turn rely on
two prominent models that map the Galactic free-electron distribution: NE2001~\cite{NE2001:CordesLazio:2002wz} and YMW16~\cite{Yao_2017}. 
Combined with direct measurements of a pulsar's radio DMs, these can be used to estimate distances to pulsars in the Milky Way. 
These models have been well-calibrated to achieve remarkable precision on distance scales of kiloparsecs (kpc) and above
with YMW16 a significant refinement of the earlier NE2001. 
Both models integrate an array of
Galactic features, including spiral arms, thin and thick disks, and localized clumps of electron density, allowing for an accurate reconstruction of the interstellar medium (ISM) structure. 
As such, for pulsar timing arrays, studies of the interstellar medium, and efforts to probe Galactic structure, the YMW16 and NE2001 models prove indispensable for finding distances across vast galactic distances. 
However, both models have some drawbacks, especially when it comes to accurately predicting distances to nearby pulsars, due to large uncertainties in estimates of the {\em local} free-electron density in the 1 kpc vicinity of the Sun. 
Upon close inspection of pulsars with distances given by radio parallaxes, these models appear not to account for severe overdensities and underdensities of electrons in this region. 
This is a point conceded by YMW16 in Ref.~\cite{Yao_2017}, where it is stated that this loss of accuracy is an unavoidable consequence of its inability to adequately model small-scale structure.

In this work, we formulate a new simplified free electron density map as an aid to finding the closest pulsar.
We calibrate this map using {\em parallax} measurements of pulsar distances within 1 kpc of Earth, as these measurements are generally known to be reliable. 
With our simplified map, we predict and list the distances of several promising pulsar candidates.
Using this method, we find some candidate nearest NSs {\em already discovered in the sky} that may be only a few tens of parsecs away.
If any of these candidate nearby NSs are confirmed by future parallax measurement, they may provide a valuable new target for testing NS properties, and espceially late-stage heating.
In the latter half of this article we illustrate this utility by studying nearby NS sensitivity for a minimal mechanism arising from dark matter: kinetic heating of NSs~\cite{NSvIR:Baryakhtar:DKHNS}, possibly augmented by self-annihilations in their interior.
We will assume two benchmark temperatures resulting from dark kinetic heating: 2500~K and 10000~K, the latter of which is possible if dark matter is clumped in microhalos~\cite{Bramante:2021dyx}.
These are below the upper bound on the coldest NS observed thus far, about 30000~K~\cite{coldestNSHST}, and may be measured by the  James Webb Space Telescope (JWST),
the Thirty Meter Telescope (TMT),
or the Extremely Large Telescope (ELT). 
Using Exposure Time Calculators available online,\footnote{www.tmt.org/etc/iris, www.eso.org/observing/etc} 
we estimate the distances to reheated NSs that would be within the sensitivity of the future TMT and ELT, for reasonable exposure times.

This paper is organized as follows.
In Section 2 we review the electron density models NE2001 and YMW16, and discuss their advantages and disadvantages.
In Section 3 we present our simplified electron density model, valid only for the solar vicinity, and use it to revise the dispersion measure distance estimates to nearby pulsars.
In Section 4 we review dark kinetic heating and forecast the observational limits obtainable by TMT and ELT to demonstrate the immediate use of our main results.
In the Appendix, we provide a detailed list of the closest pulsar candidates identified through our simplified map, including potential binary companions identified in the Gaia database, along with their revised distance estimates.

\section{Review of electron density models}
\label{sec: Electron column density models}
In pulsar astronomy, the DM is a key observable that quantifies the total column density of free electrons between the observer and a pulsar along the line of sight: 
\begin{equation}
    {\rm DM} =  \int_{0}^{L} n_e (\ell) \,d\ell~, 
    \label{eq:defn:DM}
\end{equation}
where $L$ is the distance along the line of sight and ${n_e}$ is the electron number density~\cite{2020A&A...644A.156D}. 
The lower frequency waves of a radio pulse arrive later than the higher frequency waves due to dispersion by the ISM. 

The delay in arrival times of waves with frequencies $f_1$ and $f_2$ is then given by
\begin{equation}
    \Delta t \propto {\rm DM} \left(\frac{1}{f_1^2}-\frac{1}{f_2^2}\right)~.
\label{eq:DeltatvDM}
\end{equation}

Using $\Delta t$ measurements for various lines of sight and pulse frequencies, combined with distance estimates, the spatial distribution of free electrons in the galaxy may be modeled.

This method has produced empirical electron density maps for the Milky Way, but since there may be small non-electron contributions to the DM, it has been recently advocated~\cite{DMKulkarni:2020wss} that the directly measured quantity $\mathcal{D} \equiv \Delta t (f_1^{-2} - f_2^{-2})^{-1}$  be reported by astronomers instead of the DM inferred from Eq.~\eqref{eq:DeltatvDM}.

Over the past decades several models have been developed to create and refine these maps, the most notable of which we will now briefly describe. 

\subsection{NE2001}
\label{sec:NE2001}

The NE2001 model~\cite{NE2001:CordesLazio:2002wz}, making use of measurements of pulsar DM and distances and radio-wave scattering, built upon and superseded the 1993 model of Taylor and Cordes (TC93)~\cite{TC93:1993ApJ...411..674T} for the Galactic distribution of free electrons.

The basic structure consists of three smooth components -- thick disk, thin disk, and spiral arms --, the Galactic Center, the local ISM, clumps, and voids. The NE2001 model calculates the local electron density by blending contributions from different regions of the Galaxy, each with its own distinct properties. The model begins by considering the primary electron sources, the Galactic disk and the Galactic Center, which dominate the electron density. It then accounts for the influence of the ISM and further introduces corrections for voids—regions with low electron density—and denser clumps of electrons scattered throughout the Galaxy. By assigning weights to these components, the model ensures that the distribution reflects their varying levels of influence. This weighting allows for a more nuanced and accurate representation of the electron density, especially across diverse environments within the Milky Way.

\subsection{YMW16} \label{sec:YMW16}

The YMW16 model~\cite{Yao_2017} predicts the large scale distribution of free electrons in the Galaxy, Large Magellanic Cloud (LMC), Small Magellanic Cloud (SMC), and the Intergalactic Medium (IGM). 
The Galactic part of this model follows the same basic structure as NE2001, with the addition of a four-armed spiral pattern along with a local arm, the location and form of the arms based on observations of $>$ 1800 H-II regions across the Galaxy. 
This model is then fitted to 189 independent estimates of pulsar distances that make use of parallaxes, Galactic rotation kinematics of H-I clouds with absorption features, and association with other celestial objects.

Key features in YMW16 (most of which were also part of NE2001) are: 
the Local Bubble, two regions of enhanced electron density on the periphery of the Local Bubble, the Gum Nebula, a region of enhanced electron density in the Carina arm, and a region of reduced electron density in the tangential periphery  of Sagittarius. 
One major difference between NE2001 and YMW16 is the modeling of the large scale distribution of interstellar scattering: NE2001 incorporates this effect, while YMW16 omits it. This is because numerous studies have demonstrated that interstellar scattering is often dominated by only a few regions, with significant electron density fluctuations along the path to a pulsar, making it complicated to model.
Another salient difference is that YMW16 does not incorporate clumps and voids to rectify discrepant model distances to some pulsars, in order to avoid future discrepancies for pulsars that may yet be discovered close to their lines of sight. 

DM pulsar distance estimates in the ATNF catalogue~\cite{Manchester_2005} use the YMW16 model as default. However, other distance estimates can improve accuracy. For example, distances determined by association with another object, such as the LMC or a supernova remnant, and those based on measured annual parallax (with an uncertainty less than one-third of the parallax value) are generally more reliable than distances derived from DMs. 
For the Local Bubble, there is reason to believe that there is a non-linear relationship between the DMs and the distances, (see Eq.~\eqref{eq:defn:DM}) since we expect inhomogeneities in the local ISM electron density~\cite{PulsarScintillationandLocaLBubble1998ApJ...500..262B}.

\subsection{Limitations of existing models}

Both the leading electron density maps have some limitations.
The NE2001 model has large errors in distance estimates within 1 kpc of Earth, as this model was initially designed with regard to the overall structure of the Galaxy as opposed to fine features of the local region around the Solar neighborhood. For instance, while the YMW16 model predicts a distance of 143 pc for pulsar J0536$-$7543 based on the DM, NE2001 places it at 826 pc. This discrepancy highlights the NE2001 model's broader focus on the large-scale Galactic structure, which can lead to substantial errors when estimating distances in the local region around Earth.
 Further discrepancies arise due to assumptions about local electron density variations that are ill-constrained due to the sparse pulsar DM observations available at the time of inception of this model. While the model incorporates several large scale galactic features, it doesn't give an accurate representation for smaller scale structure, which significantly impact DM interpretation for nearby pulsars.

The YMW16 model utilizes more recent observational data, including several parallax measurements for a more extensive pulsar database, but still falls short when it comes to accurately describing the local free electron density. This is due to the fact that while the model refines the large scale features, this does not always translate to a higher precision for nearby pulsars, where small scale variations are more pronounced. Both NE2001 and YMW16 contain simplifications that can introduce systematic errors for nearby pulsar distance estimates. These include assuming a smooth distribution of electrons and not accounting for small-scale clumpiness or voids in the ISM.

\section{A new local kiloparsec electron density map}
\label{sec:finding closest pulsars with new map}

\subsection{The importance of parallax measurements}

Pulsar distances are measured most robustly with parallax methods that use pulsar timing to localize their sky position~\cite{timingparallax:NANOGrav:2015oil,timingparallax:Reardon:2015kba} and/or very long baseline interferometry (VLBI)~\cite{VLBI57Pulsars:Deller:2018zxz,VLBIPulsarsReview:2014}.
A significant discrepancy between the VLBI-derived distances and those estimated from DM measurements combined with the NE2001 and YMW16 models was noted for 57 pulsars~\cite{VLBI57Pulsars:Deller:2018zxz}.
As noted in Ref.~\cite{Raj:2024kjq}, PSR J1057$-$5226 is estimated to be at 93 pc by the YMW16 model and listed as the closest pulsar in the ATNF catalogue, but is estimated to be $730 \pm 150$~pc away by the NE2001 model and 1530~pc away by the TC93 model~\cite{ClosestATNFDistsPosselt:2015bra}. 
Moreover, an analysis of the optical and x-ray spectrum~\cite{ClosestATNFOpticalXRay2010} puts PSR J1057$-$5226 at 350$\pm$150~pc. 
To our knowledge, no parallax measurement of its distance has been undertaken.

Such discrepancies arise due to inherent limitations and systematics of the DM method. This is quantified in the YMW16 model~\cite{Yao_2017} by the statement that it comes with a less-than-90\% uncertainty on 95\% of its distance estimates. Note that typical uncertainties in parallax distance estimates are 10\%–20\%,  underscoring their importance for not only pulsar distances but also the ISM distribution.
And as already noted, previous electron density models are unreliable on sub-kpc scales -- and it is exactly in the local $\mathcal{O}$(kpc) region that parallax measurements work best.
Accurate distances to neighboring pulsars are, we re-emphasize, crucial for various astrophysical studies such as that of dark matter interactions with NSs. 

\begin{figure}[h!]
  \centering
  \includegraphics[width=0.99\textwidth]{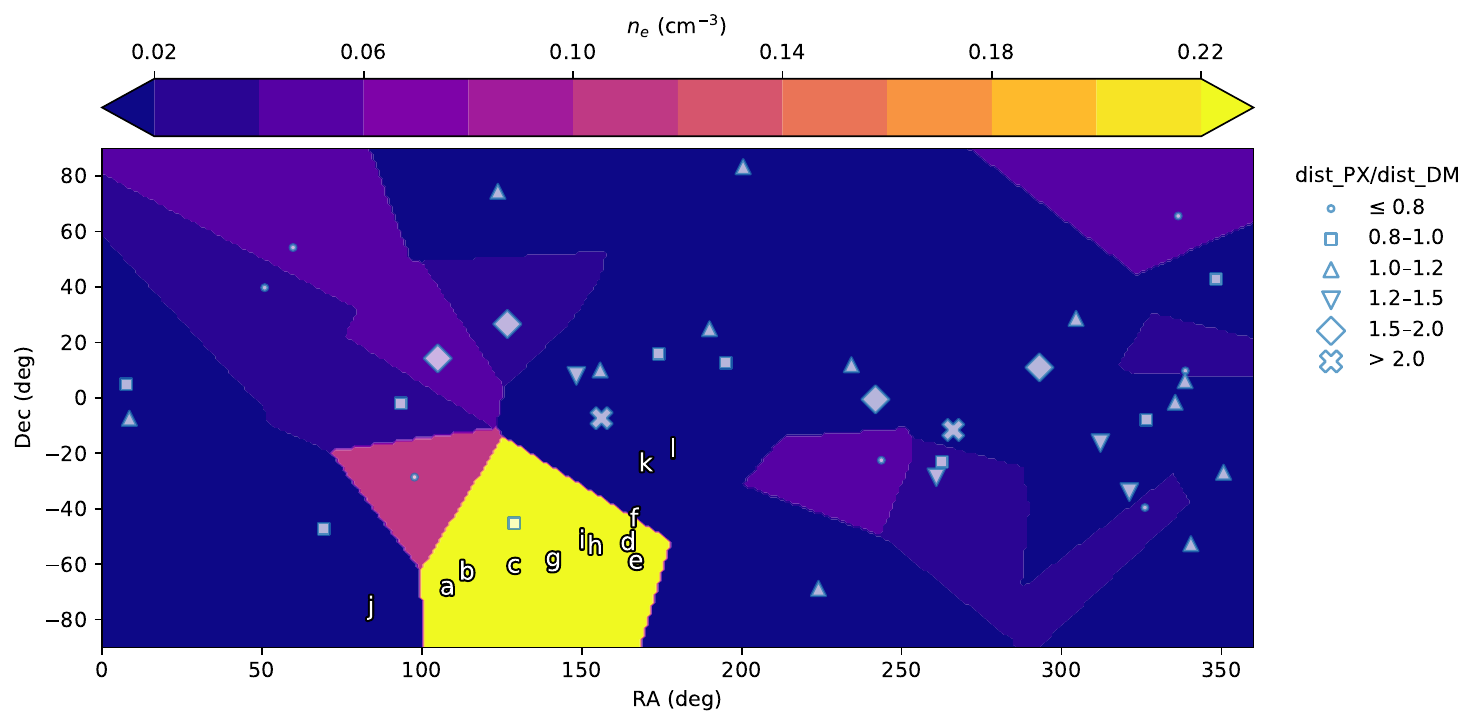}
  \vspace{1ex}

  {\small
   \setlength{\tabcolsep}{3pt}%
   \renewcommand{\arraystretch}{1.2}%
   \setlength\LTcapwidth{\textwidth}%
   \begin{tabular}{|c|l|c|c|c|c|}
     \hline
       & PSR               & DM (pc cm$^{-3}$)
                        & $\delta$DM (pc cm$^{-3}$)
                        & YMW dis (kpc)
                        & Predicted dis (kpc) \\
     \hline
     a.\ & J0711$-$6830   &  18.4096 & 0.02 & 0.106 & 0.078$\pm$0.005\\
     b.\ & J0736$-$6304   &  19.4     & -    & 0.104 & 0.082$\pm$0.005\\
     c.\ & J0834$-$60     &  20       & 6    & 0.095 & 0.084$\pm$0.030\\
     d.\ & J1057$-$5226   &  29.69    & 0.01 & 0.093 & 0.125$\pm$0.007\\
     e.\ & J1107$-$5907   &  40.75    & 0.02 & 0.115 & 0.172$\pm$0.009\\
     f.\ & J1105$-$4353   &  45       & -    & 0.127 & 0.214$\pm$0.012\\
     g.\ & J0924$-$5814   &  57  & -    & 0.107 & 0.241$\pm$0.014\\
     h.\ & J1016$-$5345   &  67  & -    & 0.117 & 0.284$\pm$0.016\\
     i.\ & J1000$-$5149   &  72.8     & 0.30 & 0.127 & 0.307$\pm$0.019\\
     j.\ & J0536$-$7543   &  18.6    & -    & 0.127 & 1.092$\pm$0.007\\
     k.\ & J1120$-$24     &   9.81    & 0.13 & 0.098 & 1.560$\pm$0.229\\
     l.\ & J1154$-$19     &  10.69    & 0.05 & 0.121 & 1.699$\pm$0.235\\
     \hline
   \end{tabular}%
  }

  \caption{%
    A new map of the free electron density (blue-yellow colorbar) for the local region, created by fitting with a zeroth-order interpolation of parallax distances to all pulsars with reported parallaxes within 1 kpc. These pulsars are shown as white symbols, and have been classified by the ratio ${\rm dis}_{\rm PX}/{\rm dis}_{\rm YMW}$. Possibly‐nearby pulsars without published parallaxes are labeled “a”–“l” in the figure; their YMW-predicted distances and $\delta DM$ (where available) are tabulated here alongside our new predicted distances from the revised $n_e$ map. Some pulsar DM measuements reported without error bars are indicated with ``-.''%
  }
  \label{fig:enter-overdensity and underdensity on neLB}
\end{figure}

\subsection{Parallax-fitted local electron distribution}\label{sec:map}

To create an electron density map for seeking the nearest pulsar, we start by compiling two datasets from the ATNF catalogue, one with pulsars for which a parallax (preferably radio parallax) is reported,  and the other with all pulsars within 1 kpc of the Sun.\footnote{In principle catalogues of pulsars discovered by FAST~\cite{pulsarcatalogueFASTCRAFTS188,pulsarcatalogueFASTGPPS637,pulsarcatFASTGPPS1Han:2021ekd,pulsarcatFASTGPPS2Zhou:2023nns,pulsarcatFASTGPPS3Su:2023vcw} and CHIME~\cite{pulsarcatalogueCHIME25,pulsarcatalogueCHIME2ndSet21} may also contain specimens within 1 kpc, as derived from the NE2001 and YMW16 models~\cite{Raj:2024kjq}. These may also be included in our analysis but we only include ATNF catalogue pulsars as their properties have been reliably verified.} Then by using the parallax distances of and the DM along the line of sight to these sub-kpc pulsars, we reconstruct the electron number density from Eq.~\eqref{eq:defn:DM}. 
Our local $n_e$ map is generated by Gaussian‐process regression using a squared‐exponential covariance kernel. The kernel’s characteristic angular length scale (i.e.\ the fitted hyperparameter) is $\sim0.4\ \mathrm{rad}$ (approximately $23^\circ$), which imposes a natural smoothing of all electron‐density features on patches of angular radius $\approx23^\circ$.  In addition, within our 1\,kpc sample the mean angular separation between neighboring pulsars is $\approx17^\circ$, further limiting the map’s intrinsic resolution.  Consequently, variations on angular scales below $\approx15$--$20^\circ$ (corresponding to physical sizes $\lesssim0.3$--$0.4\,$kpc at 1\,kpc) remain unresolved in our reconstruction.  We depict contours of this number density in Fig.~\ref{fig:enter-overdensity and underdensity on neLB} in equatorial coordinates.  (For a representation in Galactocentric coordinates, see Appendix~\ref{app:Map in galactocentric coords}.)

We have also overlaid the locations of the pulsars used as inputs;
for more details on these pulsars, see Appendix~\ref{app:pulsartables}.
Note that these sub-kpc pulsars are also part of the 189 pulsars used by YMW ~\cite{Yao_2017}, except that the fitting model is explicitly suited for larger-than-kpc scales.
To illustrate the effect of fluctuations in electron densities, we mark the ratio of their parallax distance to YMW16-based distance using different shapes as explained in the figure legend.

Finally, we also overlay (using white alphabet labels) the locations of pulsars for which there are no parallax estimates, but that which are the closest according to YMW16.
We collect these pulsars in the table below the plot, ranking them in increasing order of (the central value of) their distance as per our new map.
For each pulsar we propagate both its dispersion‐measure uncertainty \(\delta\mathrm{DM}\) and the local Gaussian‐process–predicted electron‐density uncertainty \(\delta n_e(\alpha,\delta)\) via
\begin{equation}
d \;=\;\frac{\mathrm{DM}}{n_e(\alpha,\delta)}, 
\qquad
\delta d \;=\;\sqrt{
\biggl(\frac{\delta\mathrm{DM}}{n_e(\alpha,\delta)}\biggr)^2
+
\biggl(\frac{\mathrm{DM}\,\delta n_e(\alpha,\delta)}{[\,n_e(\alpha,\delta)\,]^2}\biggr)^2
}\,,
\end{equation}
where both \(n_e\) and \(\delta n_e\) are evaluated at the pulsar’s equatorial coordinates via our regression model. This yields an uncertainty on each distance estimate that directly reflects the map‐inferred electron‐density error at that sky location.

Note that these uncertainties do not account for the Lutz-Kelker bias \cite{LUTZ1973PASP...85..573L} which arises due to the way parallax measurements are distributed and the subsequent statistical treatment of those measurements. 
The table also provides the DM and YMW16 distance estimates.
We note that Vahdat et al.\cite{Vahdat2022A&A...658A..95V} derive an X‐ray–absorption–based distance of \(\sim860\)\,pc (Section 5.2 of that work).  However, that value is not a geometric parallax but is inferred from modelling the soft X-ray hydrogen column density, and thus carries large systematic uncertainties due to inhomogeneous ISM absorption in this direction.  By calibrating exclusively to direct parallax measurements (with \(\le25\%\) uncertainty), our map predicts \(d = 78 \pm 5\)\,pc for PSR J0711–6830.  We flag this discrepancy and emphasize that future VLBI parallax observations are necessary to adjudicate between these very different estimates.

Furthermore, wherever applicable, we identify potential optical binary companions to the pulsars which may aid in the precise measurement of pulsar distances, and also provide information on dynamic interactions within such systems.
This may be the second best estimate of pulsar distances after radio parallax measurement, if the binary companion can be confirmed through other astrometric data.
The identification of an optical or infrared counterpart of a pulsar often relies on finding a star whose position coincides with the pulsar's location and exhibits characteristics of binarity, \textit{e.g.}, variability and proper motion. 
To identify binaries in the list of pulsars in the table under Fig. \ref{fig:enter-overdensity and underdensity on neLB}, we use the Infrared Science Archive (IRSA)~\cite{IRSANASA}, to take a $30$ arcsec radial field of view, and cross-reference the resulting objects with the Gaia DR3 database, checking proper motions, locations, and parallaxes; see Table~\ref{tab:Pulsars with potential binaries} in Appendix~\ref{app:pulsartables}. 
For pulsars with proper motions reported by ATNF, no promising binary candidates are found.
For the rest of the pulsars in our list, only the location and parallax can be used to ascertain the presence of binary counterparts in Gaia DR3. 
We compare these to the pulsar distances obtained from our revised map. 
This process identifies potentially four binary companions, which we summarize in Table~\ref{tab:Potential binaries} in Appendix~\ref{app:pulsartables}. 
We note that the pulsars J1000–5149, J1016–5345, J0924–5814, and J1107–5907 have very low spin-down luminosities (\(\dot E \lesssim 10^{33}\,\mathrm{erg\,s^{-1}}\)), placing them in a regime where pulsar-wind heating of any companion is effectively negligible.  In typical millisecond-pulsar binaries, a white-dwarf companion is rendered optically bright only if \(\dot E\) exceeds \(\sim10^{34}\)–\(10^{35}\,\mathrm{erg\,s^{-1}}\); below this threshold its intrinsic cooling emission (\(T_{\rm eff}\lesssim5000\)K) corresponds to \(G\)-band magnitudes \(G\gtrsim20\)–21, fainter than the \textit{Gaia}DR3 completeness limit of \(G\approx20.7\)mag (Gaia Collaboration 2022).  Observationally, low-\(\dot E\) systems rarely exhibit optical counterparts in \textit{Gaia}\cite{Tauris2014ApJ...781L..13T} .  Consequently, the absence of \textit{Gaia} detections for these four pulsars is fully consistent with their intrinsic energetics and with the properties expected of millisecond-pulsar white-dwarf companions.

The key outcome of our overall procedure is that, as seen in the table below Fig.~\ref{fig:enter-overdensity and underdensity on neLB}, our density map predicts a distance of only tens of parsecs to a non-trivial number of pulsars. 
In contrast, their DM distance per the YMW16 model is around 100$-$150 pc.
As shown in Ref.~\cite{Raj:2024kjq}, NSs at $\mathcal{O}(10)$~pc distance are ideal targets for observing late-stage reheating that imparts temperatures of as low as 2500 Kelvin.
We will illustrate this further for ELT and TMT in Sec.~\ref{sec:Exposure times with TMT and ELT}.
Thus this finding warrants follow-up observations of these pulsars to determine their parallax distance, which would be the most reliable estimate.

\section{Prospects for observing dark matter-induced neutron star reheating}

As mentioned in the Introduction, NSs may be heated in their late stage by a number of external and internal reheating mechanisms.
These include such astrophysical effects~\cite{NsvIR:otherinternalheatings:Reisenegger,NSvIR:otherinternalheatings:2022,Rotochem:FernandezReisenegger:2005cg,Rotochem:Reisenegger:2006ky,Rotochem:PetrovichReis:2009yh,Rotochem:GonzalezJimenezReis:2014iia,Rotochem:Gusakov2021:Refined,NSvIR:Hamaguchi:RotochemicalPure2019,Rotochem:GusakovReisenegger2015:CrustNSE,VortexCreep:Anderson:1975zze,VortexCreep:1984,VortexCreep:1989,VortexCreep:1993,VortexCreep:VanRiper:1994vp,VortexCreep:Larson:1998it,BfieldDecay:1992Reisenegger,CrustCrack:1971BaymPines} as rotochemical heating, vortex creep heating, crust-cracking, and magnetic field decay, as well as those induced by a new particle sector~\cite{BramanteRajCompactDark:2023djs,snowmass:ExtremeBaryakhtar:2022hbu,snowmass:Carney:2022gse,NSvIR:Raj:DKHNSOps,NSvIR:Bell2018:Inelastic,NSvIR:Bell2019:Leptophilic,NSvIR:Riverside:LeptophilicShort,NSvIR:Riverside:Leptophiliclong,NSvIR:Bell:ImprovedLepton,NSvIR:GaraniHeeck:Muophilic,NSvIR:SelfIntDM,NSvIR:Hamaguchi:RotochemicalvDM2019,NSvIR:GaraniGenoliniHambye,NSvIR:Queiroz:Spectroscopy,NSvIR:Bell:Improved,NSvIR:Bell2020improved,NSvIR:anzuiniBell2021improved,NSvIR:Marfatia:DarkBaryon,NSvIR:DasguptaGuptaRay:LightMed,NSvIR:zeng2021PNGBDM,NSvIR:Queiroz:BosonDM,NSvIR:HamaguchiEWmultiplet:2022uiq,NsvIR:HamaguchiMug-2:2022wpz,NSvIR:PseudoscaTRIUMF:2022eav,NSvIR:InelasticJoglekarYu:2023fjj,NSvIR:Hamaguchi:VortexCreepvDM2023,Bertoni:2013bsa,NSvIR:GaraniGenoliniHambye,NSvIR:BellThermalize:2023ysh,Kouvaris:2007ay,deLavallaz:2010wp,NSvIR:Baryakhtar:DKHNS,NSvIR:Raj:DKHNSOps,NSvIR:Pasta,NSvIR:IISc2022,NSvIR:hylogenenesis1:2010,NSvIR:hylogenenesis2:2011,NSvIR:coann-nucleons:JinGao2018moh,NSvIR:Marfatia:DarkBaryon,DMann:bubblenucleation:Silk2019,NSvIR:magneticBH:2020,NSvIR:clumps2021,NSvIR:tidalfifthforce:Gresham2022,NSheat:DarkBary:McKeen:2020oyr,NSheat:Mirror:McKeen:2021jbh,NSheat:Mirror:Goldman:2022brt,NSheat:Mirror:Goldman:2022rth,NSheat:Mirror:Berezhiani:2020zck,Davoudiasl:2023peu,NSheat:chainreaction:PospelovRay2024} such as dark matter capture, removal of nucleons from their Fermi seas leading to the so-called nucleon Auger effect, and baryon number-violating neutron decays.

Here we will take dark matter capture as a minimal example, giving rise to kinetic and annihilation heating, and work out the signal expectations at the forthcoming TMT and ELT.
We do this to demonstrate a concrete physics case for redrawing the electron density map to the end of seeking the nearest pulsars.

\subsection{Review of dark kinetic and annihilation heating}

For a detailed description of dark matter-induced heating of NSs, see Ref.~\cite{Bramante:2023djs}.
Here we review essential details. Dark matter particles may get captured in astrophysical objects(like NSs) if they scatter and fall into their gravitational potential.
The total mass rate of the dark matter going through a NS of mass $M$ and radius $R$ is 
\begin{eqnarray}
\nonumber    \Dot{M}_\chi &=& \pi b_{\rm max}^2 \rho_{\chi} v_{\chi}~, \\
  b_{\rm max}&=&\left(\frac{2GMR}{v_{\chi}^2}\right)^{1/2}\left(1-\frac{2GM}{R}\right)^{-1/2} = \gamma R\frac{v_{\rm esc}}{v_{\chi}}~,
\end{eqnarray}
where $b_{\rm max}$ is the maximum impact parameter,  $v_{\rm esc} = \sqrt{2 G M/R}$ is the escape speed at the NS surface, and $\rho_{\chi}$ and $v_\chi$ are respectively the ambient dark matter density and halo dark matter speed, which we take as 0.42 GeV $\text{cm}^{-3}$ and 230 km $\text{s}^{-1}$ respectively~\cite{Pato:2015dua}. 

The rate of kinetic energy deposition is then 
\begin{eqnarray}
\nonumber    \Dot{E_k} &=& \frac{\Dot{M}_\chi}{m_\chi} (\gamma - 1) f~,\\
  f &=& \min \left[1,\frac{\sigma_{\chi \rm T}}{\sigma_{\rm crit}}\right]~,
\end{eqnarray}
where $f$ is the fraction of incident dark matter particles that capture,
with $\sigma_{\chi \rm T}$ the cross section for scattering with some target (nucleon, lepton, etc.)
and $\sigma_{\rm crit}$ the cross section above which the NS becomes optically thick to the infalling dark matter.
Absent Pauli-blocking and multiscatter effects, this is the NS's $\mathcal{O}(10^{-45})$~cm$^{-2}$ geometric cross section.

NSs have internal temperatures of $10^{11}$~K when formed and cool down through neutrino and photon emission, the latter dominating after $\mathcal{O}(10^5)$ yr. Until about $10^7$ yr an insulating envelope keeps the internal temperature larger than the surface temperature, but beyond this timescale it becomes too thin and the two temperatures become equal, with a value $\leq \mathcal{O}(10^3)$~K.   
Under equilibrium between dark kinetic heating and passive cooling, 
\begin{equation}
  \dot{E}_k = L_{\rm NS} = 4\pi \sigma_{\rm SB} R^2 T_s^4~,
\end{equation}
where $\sigma_{\rm SB}$ is the Stefan-Boltzmann constant, and $T_s$ is the NS surface blackbody temperature, andfor a distant observer, $T_\infty = T_s/\gamma$.
Potential NS reheating from astrophysical effects and ISM accretion is in most cases unlikely; see Ref.~\cite{BramanteRajCompactDark:2023djs}.

If the captured dark matter thermalizes with the NS rapidly enough~\cite{Bertoni:2013bsa,NSvIR:GaraniGuptaRaj:Thermalizn} and self-annihilates efficiently into particles that are trapped in the star, higher heating luminosities are attained.
In all, accretion of dark matter that is homogeneously distributed in the halo would give rise to NS surface temperatures of at best around 2500 K near the solar vicinity. 
Capture of dark matter clumped in overdensities such as microhalos, particularly in models where the dark matter has self-interactions and could thus undergo Bondi accretion, could result in heating-induced NS temperatures of $\mathcal{O}(10^4)$~K~\cite{NSvIR:clumps2021}.
Similar temperatures could be achieved in the presence of long-range interactions between the infalling dark matter and the NS baryons~\cite{NSvIR:Gresham:2022biw}.
Motivated by these considerations, we will use reheated NS temperatures of 2500 K, 6000 K, and 10000 K as benchmarks for our treatment of telescope sensitivities.

\subsection{Measuring neutron star temperatures with TMT and ELT}
\label{sec:Exposure times with TMT and ELT}

The currently operational Near Infrared Camera (NIRCam) on the James Webb Space Telescope (JWST), along with the upcoming Multi-AO Imaging Camera for Deep Observations (MICADO) on the Extremely Large Telescope (ELT) and the future InfraRed Imaging Spectrograph (IRIS) on the Thirty Meter Telescope (TMT) can image in infrared to far-optical wavelengths.
These correspond to peak blackbody temperatures of $1300-4300$~K, making these imaging instruments suitable for detecting NS reheating. 
As tabulated in Ref.~\cite{Raj:2024kjq}, with $10^5-10^6$~s of exposure, these instruments could detect $\mathcal{O}(10^3)$~K NSs that are within $\mathcal{O}(10)$~pc and $\mathcal{O}(10^4)$~K NSs within $\mathcal{O}(10^2-10^3)$~pc.
Here we carry out a similar calculation for the forthcoming ELT-MICADO and TMT-IRIS in order to make our study self-contained.

Assuming the NS to be a black body, the spectral flux density is given by

\begin{equation}\label{eq:specfluxdensity}
 f_\nu = \pi \frac{2 h \nu^3}{c^2} \frac{1}{\exp(h\nu/k T_\infty)-1} \bigg(\frac{R \gamma}{d}\bigg)^2,
\end{equation}
often re-expressed in terms of the AB magnitude,
\begin{equation}
    m_{\rm AB}=-2.5 \log_{10}\left(\frac{f_\nu}{3631\: {\rm Jy}}\right)~,
    \label{eq:AB}
\end{equation}
where $R$ and $d$ are the radius and distance of the NS respectively, and the factor $R\gamma /d$ is the angle subtended by the NS at a distant observer. 
We have neglected extinction factors along the line of sight, which would introduce uncertainties of at worst 10\%, comparable to or smaller than distance uncertainties; see Ref.~\cite{Raj:2024kjq} for a detailed discussion. 

\begin{figure}[htbp]
  \centering
  \includegraphics[width=0.44\textwidth]{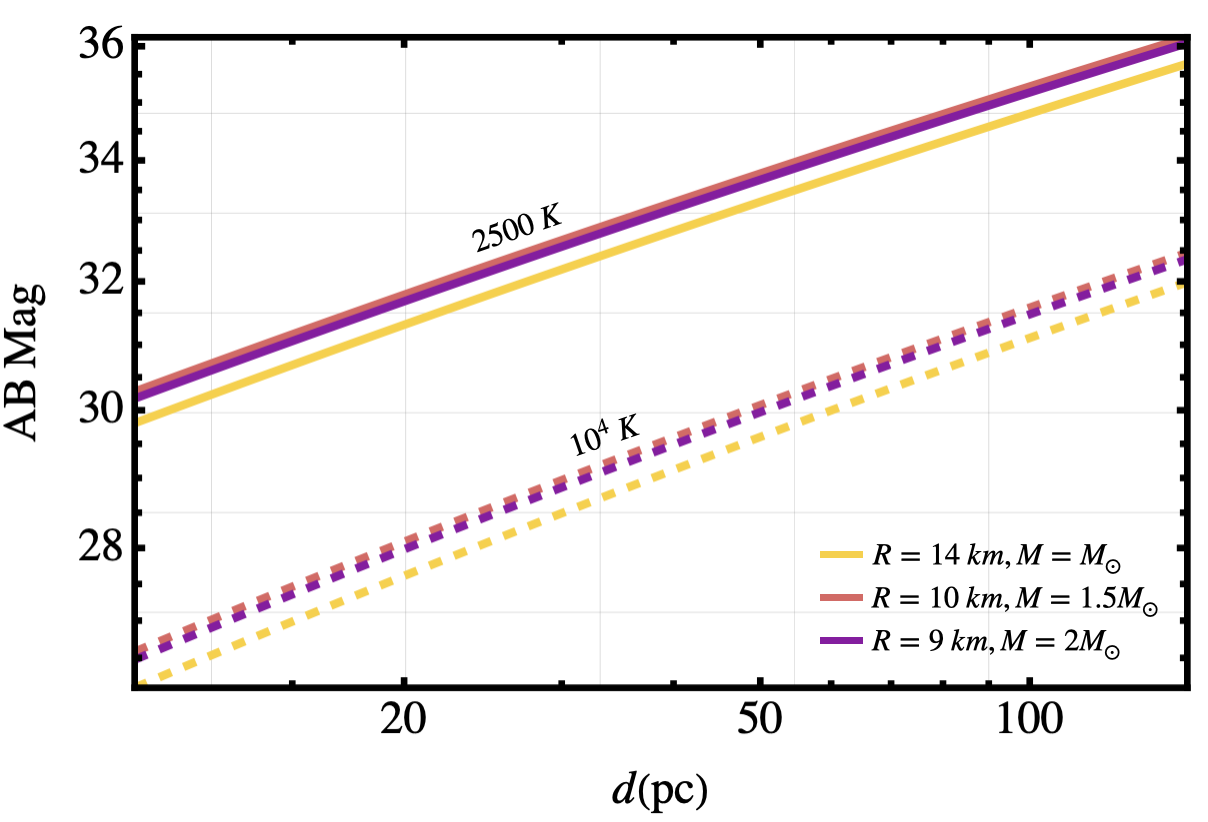}
  \includegraphics[width=0.54\textwidth]{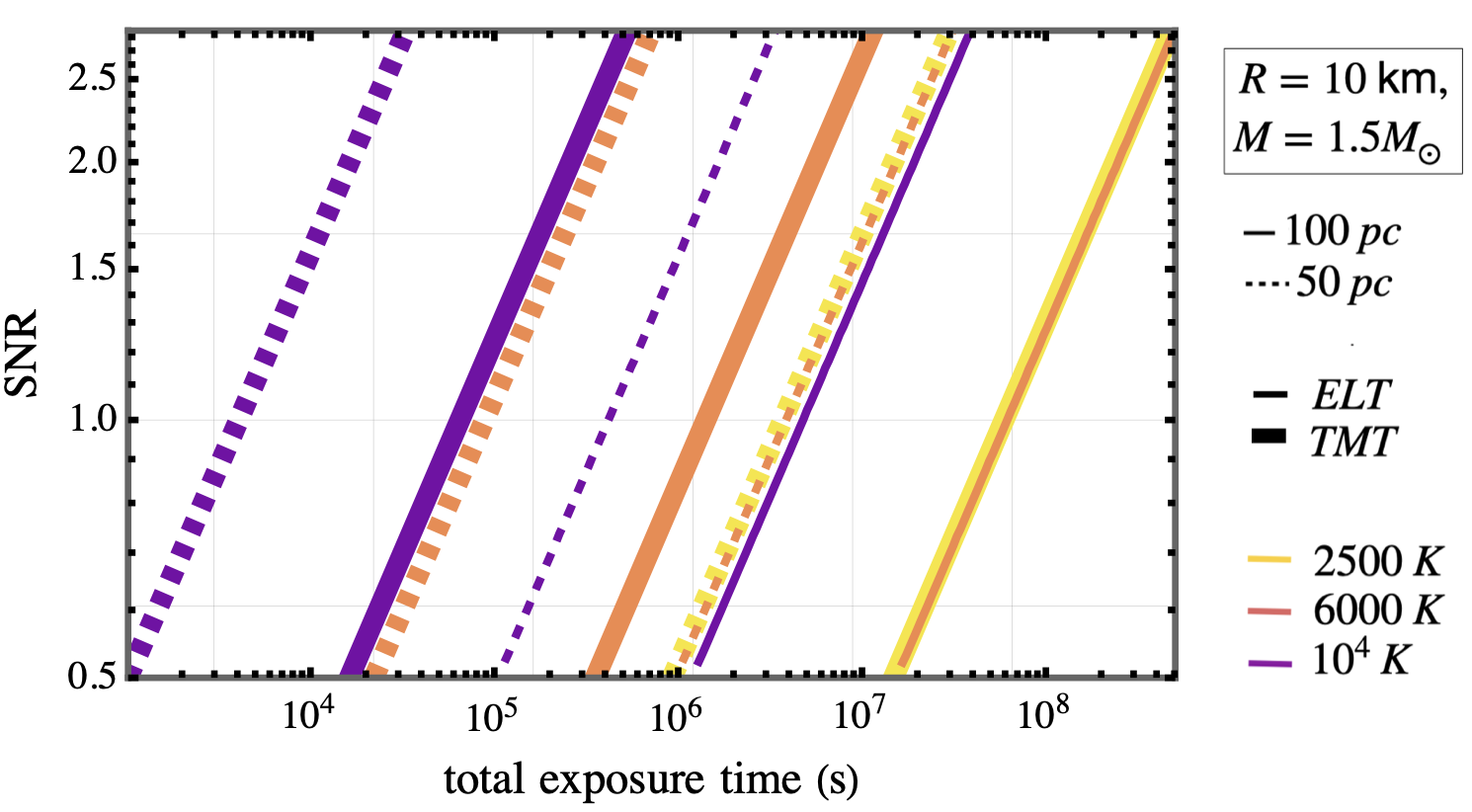}
  \caption{%
    {\bf Left:} AB magnitude derived for neutron stars heated maximally by kinetic and annihilation heating of dark matter to reach a temperature of  2500~K (solid lines) and $10^4$ K (dashed lines) for different radius and mass profiles. 
     {\bf Right:} Predicted signal‐to‐noise ratio (SNR) as a function of total exposure time for neutron stars at 50 (dashed lines) and 100 (solid lines)\,pc, computed with the MICADO ETC for the ELT and the IRIS ETC for the TMT.  For ELT–MICADO we choose the B band for $T_{\rm eff}=6000\,$K and the U band for $T_{\rm eff}=10^4\,$K, so as to sample the blackbody peak in each case.  For TMT–IRIS, we use the Y-band filter—although it does not fall at the peak of the $6000\,$K or $10^4\,$K spectra, its wide bandwidth and high throughput still yield strong photon counts and competitive sensitivities. All calculations assume a canonical neutron‐star radius of $R=10\,$km and mass $M=1.5\,M_\odot$. Note that exposures much longer than a few 10$^7$~s ($\sim$ yr) may be unrealistic, although this exposure time is comparable to prior ultra deep field observations \cite{Illingworth:2013pda}.%
  }
  \label{fig:SFD for 2500 K and 10k K and AB mag}
\end{figure}
 
Neglecting dithering and read-out pattern effects, the signal-to-noise ratio (SNR) in the background-dominated regime may be approximated as
\begin{equation}
    {\rm SNR} = \frac{\Phi_{\rm sig} A_{\rm SNR} t_{\rm exp}}{\sqrt{(\Phi_{\rm bg}A_{\rm SNR}+\Gamma_{\rm noise})t_{\rm exp}}},
\end{equation}
where
$\Phi_{\rm sig}$ and $\Phi_{\rm bg}$ are signal and background fluxes, $A_{\rm SNR}$ is the SNR reference area in the detector,  and $\Gamma_{\rm noise}$ is the non-sky noise rate.
Thus we expect $t_{\rm exp} \propto$~(SNR)$^2$, and using this in Eq.~\eqref{eq:specfluxdensity}, we have the scaling $t_{\rm exp} \propto~d^4$.
In the left panel of Fig.~\ref{fig:SFD for 2500 K and 10k K and AB mag} we plot the AB magnitudes as a function of neutron‐star distance for surface temperatures of 2500 K (solid lines) and 10000 K (dashed lines), over a range of mass–radius configurations.  As expected from Eqs.~\eqref{eq:specfluxdensity} and \eqref{eq:AB}, $m_{\rm AB}\propto\log d$, and more massive or larger stars appear brighter.
In the right panel we plot the signal–to–noise ratio versus total exposure time for neutron stars at fiducial distances of $50\,$pc and $100\,$pc.  For ELT–MICADO we display only the $T = 6000\,$K (B-band) and $T = 10^{4}\,$K (U-band) curves, since at J-band the required exposure times exceed $10^{8}\,$s—well beyond practical limits.  For TMT–IRIS we show all three temperature benchmarks ($T = 2500\,$K, $6000\,$K, and $10^{4}\,$K) using the Y-band filter.  Although Y-band does not coincide exactly with the spectral peaks of the higher‐temperature cases, its broad passband and high throughput still yield robust SNR estimates.  This choice provides a conservative, proof-of-concept sensitivity in the absence of B or U filters in the current TMT–IRIS ETC.
For the ELT, we set the source geometry to be a point source with the source spectral type corresponding to the temperature profile being considered, with a S/N reference area of $1\times1$ pixels at the Paranal observation site at $2635$ m, with a telescope diameter of $39$ m, and we use the default background model. Furthermore, we use typical values for air-mass of $1.50$, pixel scale of $50\:{\rm mas/pixel}$ and seeing limited (FWHM$=0".8$) adaptive optics mode. Similarly for TMT, we set the source geometry to be a point source, at a zenith angle of $0$ in good weather conditions, under the Imager mode configuration. 
In this figure we see the scaling SNR $\propto \sqrt{t_{\rm exp}} \propto d^2$. 
We also find the results consistent with Ref.~\cite{Raj:2024kjq}.

\section{Conclusion}
\label{sec: Conclusion}

The aim of this study is to initiate a robust search for neutron stars nearest to Earth. 
A close enough NS would be a target for observing, among other things, novel late-time reheating mechanisms including from the capture of dark matter.
To this end, we have attempted to refine the electron density map in the solar vicinity, in turn improving the accuracy of dispersion measure-driven distance estimates for pulsars. 

In particular, using parallax measurements of pulsar distances (which are unrelated to DM-based distance estimates) we constructed a simplified electron density map.
This revised map allowed us to pinpoint promising nearby pulsar candidates that are ideal for follow-up parallax observations, as summarized in the table in Figure \ref{fig:enter-overdensity and underdensity on neLB}. 
 We also investigated the possibility of finding binary stellar counterparts for these pulsars, summarized in Table \ref{tab:Pulsars with potential binaries}.

We believe our map may offer a more accurate depiction of pulsar locations within the $1$ kpc vicinity than the earlier NE2001 and YMW16 models, which had focused on large-scale galactic DMs, and included a template fit to assumed locations of electron over-densities and under-densities.
In contrast, by only concentrating on the local kpc, we provide a new estimate of the free electron density. Looking forward, more accurate estimates of distances to nearby NSs such as those undertaken in this study are an essential step to begin fathoming some longstanding puzzles of fundamental physics.

\acknowledgments

We thank M. A. Krishnakumar, Joseph Lazio, and Avinash Kumar Paladi for helpful discussions and correspondence. 
This work was supported by the Arthur B. McDonald Canadian Astroparticle Physics Research Institute, the Natural Sciences and Engineering Research Council of Canada (NSERC), and the Canada Foundation for Innovation. Research at Perimeter Institute is supported by the Government of Canada through the Department of Innovation, Science, and Economic Development, and by the Province of Ontario.
This research has made use of the NASA/IPAC Infrared Science Archive (IRSA), which is funded by the National Aeronautics and Space Administration and operated by the California Institute of Technology, the NASA/IPAC Extragalactic Database (NED),
which is operated by the Jet Propulsion Laboratory, California Institute of Technology,
under contract with the National Aeronautics and Space Administration. LS is supported by the National SKA Program of China (2020SKA0120300) and the Max Planck Partner Group Program funded by the Max Planck Society.
\newpage

\appendix
\section*{Appendices}
\addcontentsline{toc}{section}{Appendices}
\renewcommand{\thesubsection}{\Alph{subsection}}

\subsection{Pulsar properties}
\label{app:pulsartables}

In this section we collect details of pulsars used in this work.
In Table~\ref{tab:for_ne_map} are the pulsars used to construct our electron density map in Fig.~\ref{fig:enter-overdensity and underdensity on neLB} and Fig.~\ref{fig:1.1kpc}.
In Table~\ref{tab:Pulsars with potential binaries} are the pulsars for which we have estimated DM-based distances using our revised map. 
These are the same pulsars as those tabulated below Fig.~\ref{fig:enter-overdensity and underdensity on neLB}, but now we provide further details such as their proper motion.
This information is used to identify potential binary companions, listed in Table~\ref{tab:Potential binaries}.
\begingroup
  \small                        
  \setlength{\tabcolsep}{3pt}
  \renewcommand{\arraystretch}{1.2}
  \setlength\LTcapwidth{\textwidth}
\begin{longtable}{|c|c|c|c|c|c|c|}
\caption{Pulsars used to construct our local electron density map in Fig.~\ref{fig:enter-overdensity and underdensity on neLB}(and Fig.~\ref{fig:1.1kpc}), along with their reported parallax, equatorial co‐ordinate positions, dispersion measure, parallax distance and YMW‐based distance.}
\label{tab:for_ne_map}\\
\hline
\textbf{PSR} & \textbf{PX (mas)} & \textbf{RAJD (deg)} & \textbf{DECJD (deg)} & \textbf{DM (cm$^{-3}$ pc)} & \textbf{dis$_{\rm PX}$ (kpc)} & \textbf{dis$_{\rm YMW}$ (kpc)} \\
\hline
\endfirsthead

\multicolumn{7}{c}{{\bfseries \tablename\ \thetable{} -- continued from previous page}} \\
\hline
\textbf{PSR} & \textbf{PX (mas)} & \textbf{RAJD (deg)} & \textbf{DECJD (deg)} & \textbf{DM (cm$^{-3}$pc)} & \textbf{dis$_{\rm PX}$ (kpc)} & \textbf{dis$_{\rm YMW}$ (kpc)} \\
\hline
\endhead

\hline \multicolumn{7}{r}{{Continued on next page}} \\ \hline
\endfoot

\hline
\endlastfoot

J0030+0451  & 2.910 $\pm$ 0.180 &  7.61 &   4.86 &  4.33294 $\pm$ 0.00011 & 0.344 & 0.345 \\
J0034-0721  & 0.930 $\pm$ 0.080 &  8.54 &  –7.36 & 10.92200 $\pm$ 0.00600 & 1.075 & 0.996 \\
J0323+3944  & 1.051 $\pm$ 0.040 & 50.86 &  39.75 & 26.18975 $\pm$ 0.00093 & 0.951 & 1.197 \\
J0358+5413  & 0.910 $\pm$ 0.160 & 59.72 &  54.22 & 57.14200 $\pm$ 0.00030 & 1.099 & 1.594 \\
J0437-4715  & 6.430 $\pm$ 0.040 & 69.32 & –47.25 &  2.64476 $\pm$ 0.00007 & 0.156 & 0.156 \\
J0613-0200  & 1.010 $\pm$ 0.090 & 93.43 &  –2.01 & 38.77500 $\pm$ 0.00050 & 0.990 & 1.024 \\
J0630-2834  & 3.009 $\pm$ 0.409 & 97.71 & –28.58 & 34.42500 $\pm$ 0.00100 & 0.332 & 2.069 \\
J0659+1414  & 3.470 $\pm$ 0.360 &1 04.95 &  14.24 & 13.94000 $\pm$ 0.09000 & 0.288 & 0.159 \\
J0737-3039A$/$B & 0.870 $\pm$ 0.140 & 114.46 & -30.66 & 48.91600 $\pm$ 0.00200 & 1.149	& 1.105\\
J0814+7429  & 2.310 $\pm$ 0.040 &123.75 &  74.48 &  5.75066 $\pm$ 0.00048 & 0.433 & 0.368 \\
J0826+2637  & 2.010 $\pm$ 0.013 &126.71 &  26.62 & 19.47633 $\pm$ 0.00018 & 0.498 & 0.314 \\
J0835-4510  & 3.500 $\pm$ 0.200 &128.84 & –45.18 & 67.77100 $\pm$ 0.00900 & 0.286 & 0.328 \\
J0953+0755  & 3.820 $\pm$ 0.070 &148.29 &   7.93 &  2.96927 $\pm$ 0.00008 & 0.262 & 0.187 \\
J1022+1001  & 1.160 $\pm$ 0.080 &155.74 &  10.03 & 10.25500 $\pm$ 0.00080 & 0.862 & 0.834 \\
J1024-0719  & 0.970 $\pm$ 0.130 &156.16 &  –7.32 &  6.48640 $\pm$ 0.00020 & 1.031 & 0.382 \\
J1136+1551  & 2.687 $\pm$ 0.018 &174.01 &  15.85 &  4.84066 $\pm$ 0.00034 & 0.372 & 0.414 \\
J1239+2453  & 1.160 $\pm$ 0.080 &189.92 &  24.90 &  9.25159 $\pm$ 0.00053 & 0.862 & 0.827 \\
J1300+1240  & 1.410 $\pm$ 0.080 &195.01 &  12.68 & 10.16550 $\pm$ 0.00003 & 0.709 & 0.877 \\
J1321+8323  & 0.968 $\pm$ 0.140 &200.44 &  83.39 & 13.31624 $\pm$ 0.00076 & 1.033 & 0.977 \\
J1456-6843  & 2.200 $\pm$ 0.300 &224.00 & –68.73 &  8.61300 $\pm$ 0.00400 & 0.455 & 0.436 \\
J1537+1155  & 1.070 $\pm$ 0.090 &234.29 &  11.93 & 11.61944 $\pm$ 0.00002 & 0.935 & 0.877 \\
J1607-0032  & 0.934 $\pm$ 0.047 &241.80 &  –0.54 & 10.68230 $\pm$ 0.00010 & 1.071 & 0.680 \\
J1614-2230  & 1.300 $\pm$ 0.200 &243.65 & –22.51 & 34.48500 $\pm$ 0.00030 & 0.769 & 1.394 \\
J1723-2837  & 1.077 $\pm$ 0.054 &260.85 & –28.63 & 19.68800 $\pm$ 0.01300 & 0.929 & 0.720 \\
J1730-2304  & 2.300 $\pm$ 0.200 &262.59 & –23.08 &  9.62570 $\pm$ 0.00050 & 0.435 & 0.512 \\
J1744-1134  & 2.610 $\pm$ 0.090 &266.12 & –11.58 &  3.13849 $\pm$ 0.00012 & 0.383 & 0.148 \\
J1932+1059  & 2.770 $\pm$ 0.070 &293.06 &  10.99 &  3.18321 $\pm$ 0.00016 & 0.361 & 0.229 \\
J2018+2839  & 1.030 $\pm$ 0.100 &304.52 &  28.67 & 14.19770 $\pm$ 0.00060 & 0.971 & 0.959 \\
J2048-1616  & 1.050 $\pm$ 0.030 &312.15 & –16.28 & 11.45600 $\pm$ 0.00500 & 0.952 & 0.775 \\
J2124-3358  & 2.300 $\pm$ 0.200 &321.18 & –33.98 &  4.59540 $\pm$ 0.00030 & 0.435 & 0.360 \\
J2144-3933  & 6.051 $\pm$ 0.560 &326.05 & –39.57 &  3.35000 $\pm$ 0.01000 & 0.165 & 0.289 \\
J2145-0750  & 1.600 $\pm$ 0.300 &326.46 &  –7.84 &  9.00080 $\pm$ 0.00130 & 0.625 & 0.693 \\
J2222-0137  & 3.723 $\pm$ 0.014 &335.52 &  –1.62 &  3.28260 $\pm$ 0.00020 & 0.269 & 0.267 \\
J2225+6535  & 1.203 $\pm$ 0.204 &336.47 &  65.59 & 36.44362 $\pm$ 0.00051 & 0.831 & 1.881 \\
J2234+0611  & 1.050 $\pm$ 0.040 &338.60 &   6.19 & 10.76700 $\pm$ 0.00020 & 0.952 & 0.854 \\
J2234+0944  & 1.300 $\pm$ 0.400 &338.70 &   9.74 & 17.83230 $\pm$ 0.00020 & 0.769 & 1.587 \\
J2241-5236  & 0.960 $\pm$ 0.040 &340.43 & –52.61 & 11.41126 $\pm$ 0.00003 & 1.042 & 0.963 \\
J2313+4253  & 0.930 $\pm$ 0.070 &348.29 &  42.89 & 17.27693 $\pm$ 0.00033 & 1.075 & 1.108 \\
J2322-2650  & 1.300 $\pm$ 0.200 &350.64 & –26.85 &  6.14906 $\pm$ 0.00013 & 0.769 & 0.760 \\

\end{longtable}
\endgroup

\begin{table}[h!]
\centering
\small
\setlength{\tabcolsep}{3pt}
\renewcommand{\arraystretch}{1.2}
\setlength\LTcapwidth{\textwidth}
  \begin{tabular}{|c|c|c|c|c|c|c|c|l|}
    \hline
    & \textbf{PSR} & \textbf{RA} & \textbf{DEC} & \textbf{Pred.\ dist.} & \textbf{PM RA} & \textbf{PM DEC} & \textbf{PM tot} & \textbf{Potential binary} \\
    &        & (deg)       & (deg)        & (kpc)& (mas/yr)      & (mas/yr)       & (mas/yr)       &                \\\hline
    a.& J0711$-$6830 & 107.97 & $-68.51$ & $  0.078\pm 0.005$& $-15.56\pm 9$ & $ 14.18\pm 9$ & $ 21.05\pm 9$ &                   \\
    b.& J0736$-$6304 & 114.08 & $-63.07$ & $  0.082\pm 0.005$&  ---          & ---           & ---           &                   \\
    c.& J0834$-$60   & 128.71 & $-60.58$ & $  0.084\pm 0.030$&  ---          & ---           & ---           &                   \\
    d.& J1057$-$5226 & 164.49 & $-52.44$ & $ 0.125\pm 0.007$& $ 47.5\pm 7$  & $ -8.7\pm 7$  & $ 48.3\pm 7$  &                   \\
    e.& J1107$-$5907 & 166.89 & $-59.12$ & $ 0.172\pm 0.009$&  ---          & ---           & ---           & Gaia DR3 5338633770476031360 \\
    f.& J1105$-$4353 & 166.35 & $-43.88$ & $ 0.189\pm 0.011$&  ---          & ---           & ---           &                   \\
    g.& J0924$-$5814 & 141.13 & $-58.23$ & $ 0.241\pm 0.014$&  ---          & ---           & ---           & Gaia DR3 5306447697839727360 \\
    h.& J1016$-$5345 & 154.13 & $-53.75$ & $ 0.284\pm0.016$&  ---          & ---           & ---           & Gaia DR3 5356512791576521984 \\
    i.& J1000$-$5149 & 150.12 & $-51.83$ & $ 0.307\pm0.019$&  ---          & ---           & ---           & Gaia DR3 5405166971370873216 \\
    j.& J0536$-$7543 &  84.13 & $-75.73$ & $ 1.092\pm0.007$& $  3\pm40$    & $ 65\pm 8$    & $ 65\pm 9$    &                   \\
    k.& J1120$-$24   & 170.00 & $ -24.00$ & $  0.824\pm 0.068$&  ---          & ---           & ---           &                   \\
    l.& J1154$-$19   & 178.50 & $ -19.00$ & $  0.822\pm 0.009$&  ---          & ---           & ---           &                   \\
 \hline
  \end{tabular}%
\caption{Pulsars (a--l) referenced in Figs.~\ref{fig:enter-overdensity and underdensity on neLB} $\&$~\ref{fig:1.1kpc}, with their predicted distances from our $n_e$ map(Fig.~\ref{fig:1.1kpc}, proper motions, and any Gaia DR3 companion.}
\label{tab:Pulsars with potential binaries}
\end{table}

\begin{table}[h!]
\centering
\small
\setlength{\tabcolsep}{3pt}
\renewcommand{\arraystretch}{1.2}
\setlength\LTcapwidth{\columnwidth}
\resizebox{\textwidth}{!}{%
  \begin{tabular}{|l|c|c|c|c|c|c|c|c|c|}
    \hline
    \textbf{Potential binary} & \textbf{RA} & \textbf{DEC} & \textbf{PM RA} & \textbf{PM DEC} & \textbf{PM tot} & \textbf{PX} & \textbf{dist$_{\rm PX}$} & \textbf{Assoc.\ PSR} & \textbf{Pred.\ dist.} \\
                               & (deg)      & (deg)       & (mas/yr)      & (mas/yr)       & (mas/yr)       & (mas)      & (kpc)                   &                   & (kpc)                \\ \hline
    Gaia DR3 5338633770476031360 & 166.91 & $-59.12$ & $-4.25\pm1.24$ & $ 2.35\pm1.10$ & $ 4.86$ & $3.13\pm1.46$ & 0.319 & J1107$-$5907 & $0.172\pm 0.009$ \\
    Gaia DR3 5306447697839727360 & 141.14 & $-58.23$ & $-3.56\pm0.92$ & $ 3.39\pm1.21$ & $ 4.91$ & $1.54\pm1.03$ & 0.649 & J0924$-$5814 & $0.241\pm 0.014$ \\
    Gaia DR3 5356512791576521984 & 154.14 & $-53.75$ & $-5.03\pm0.94$ & $ 2.17\pm0.93$ & $ 5.48$ & $1.72\pm0.73$ & 0.581 & J1016$-$5345 & $0.284\pm0.016$ \\
    Gaia DR3 5405166971370873216 & 150.12 & $-51.83$ & $-5.09\pm1.08$ & $ 4.80\pm0.95$ & $ 7.00$ & $1.73\pm0.81$ & 0.578 & J1000$-$5149 & $0.307\pm0.019$ \\ \hline
  \end{tabular}%
}
\caption{Gaia DR3 counterparts to the pulsars in Table~\ref{tab:Pulsars with potential binaries}, with their parallaxes and parallax distances, compared to the predicted distances from our $n_e$ map(Fig.~\ref{fig:1.1kpc}.)}
\label{tab:Potential binaries}
\end{table}

\newpage
\subsection{Modified electron density map in galactocentric representation}
\label{app:Map in galactocentric coords}

In this section, we re-create Fig.~\ref{fig:enter-overdensity and underdensity on neLB}, with the region of interest extended to $1.1$ kpc.

\begin{figure}[h!]
    \centering
     \includegraphics[width=0.9\textwidth]{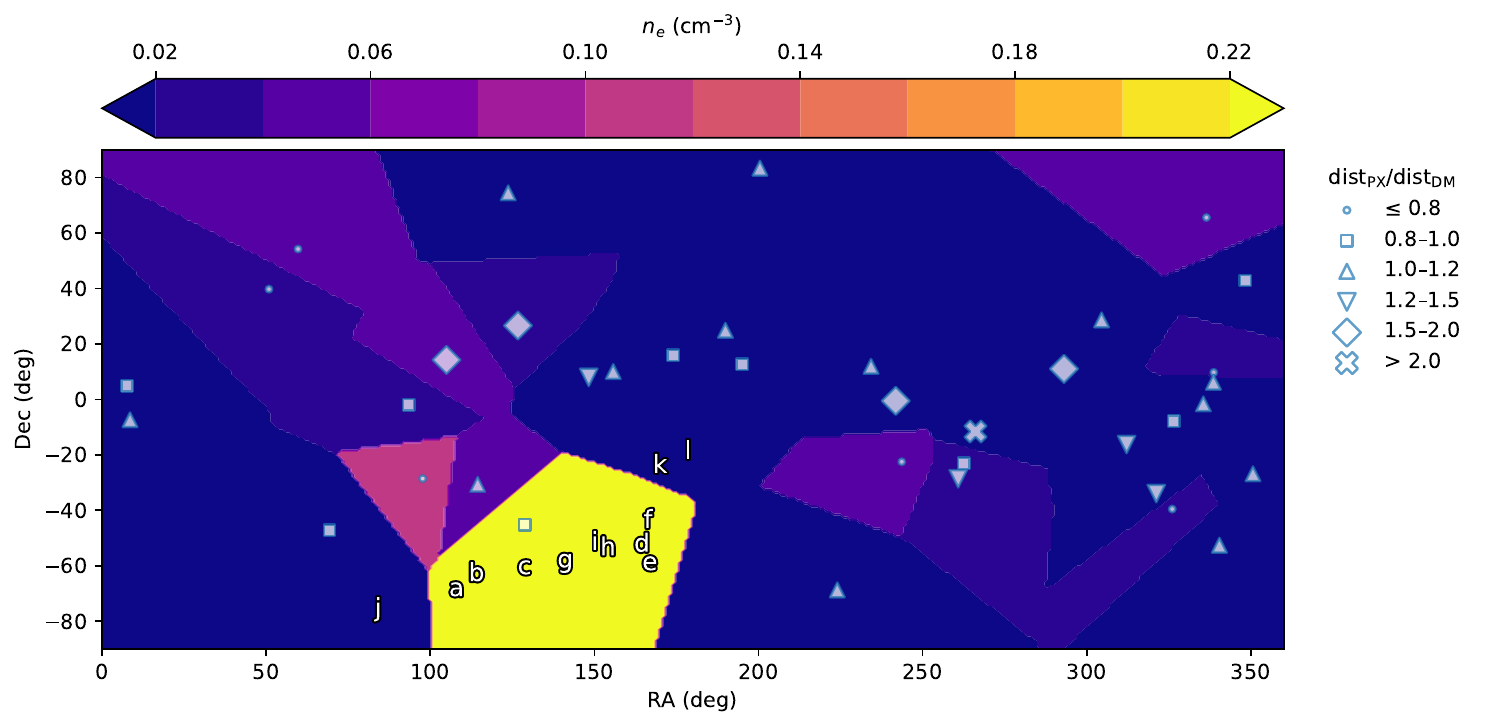}
    \centering
\begin{tabular}{|c|l|c|c|c|c|}
\hline
 & \text{PSR} & \text{DM} (${\rm pc}\:{\rm cm}^{-3})$ & \text{$\delta$DM} (${\rm pc}\:{\rm cm}^{-3})$ & \text{YMW dis (kpc)} & \text{Predicted dis (kpc)}\\
\hline
a. & J0711$-$6830 &  18.4096 & 0.02 & 0.106 & 0.078 $\pm$ 0.005\\
b. & J0736$-$6304 &  19.4 & - & 0.104 & 0.082 $\pm$ 0.005\\
c. & J0834$-$60 &  20 & 6 & 0.095 & 0.084 $\pm$ 0.030\\
d. & J1057$-$5226 & 29.69 & 0.01 & 0.093 & 0.125 $\pm$ 0.007 \\
e. & J1107$-$5907 & 40.75 & 0.02 & 0.115 & 0.172 $\pm$ 0.009\\
f. & J1105$-$4353 &  45 & - & 0.127 & 0.189 $\pm$ 0.011\\
g. & J0924$-$5814 &  57 & - & 0.107 & 0.241 $\pm$ 0.014\\
h. & J1016$-$5345 &  67 & - & 0.117 & 0.284 $\pm$ 0.016\\
i. & J1000$-$5149 &  72.8 & 0.30 & 0.127 & 0.307 $\pm$ 0.019\\
j. & J0536$-$7543 &  18.6 & - & 0.127 & 1.092 $\pm$ 0.007\\
k. & J1120$-$24 &  9.81 & 0.13 & 0.098 & 0.824	$\pm$ 0.068\\
l. & J1154$-$19 &  10.69  & 0.05 & 0.121 & 0.822 $\pm$ 0.009\\
\hline
\end{tabular}
    \caption{Map of the free electron density (blue to yellow color bar) for the extended region of 1.1 kpc, created by fitting with a zeroth-order interpolation of parallax distances to all pulsars with reported parallax measurements that are within 1.1 kpc of Earth. These pulsars are displayed as white symbols, and have been classified according to the ratio of their distance estimates from parallax ($dis_{PX}$) and from the YMW16 model ($dis_{YMW}$). Possibly nearby pulsars that do not yet have parallax measurements made on them are tabulated and labeled “a”$-$“l” in the figure. Their YMW16 predicted distances and $\delta DM$ (where available) are listed in the table alongside new predicted distances derived from our revised electron density map.
    Discrepancies with Table~\ref{fig:enter-overdensity and underdensity on neLB} are seen to be conspicuous near the 1 kpc scale.}
    \label{fig:1.1kpc}
\end{figure}

\begin{figure}[htbp]
    \centering
    \includegraphics[width=0.99\linewidth]{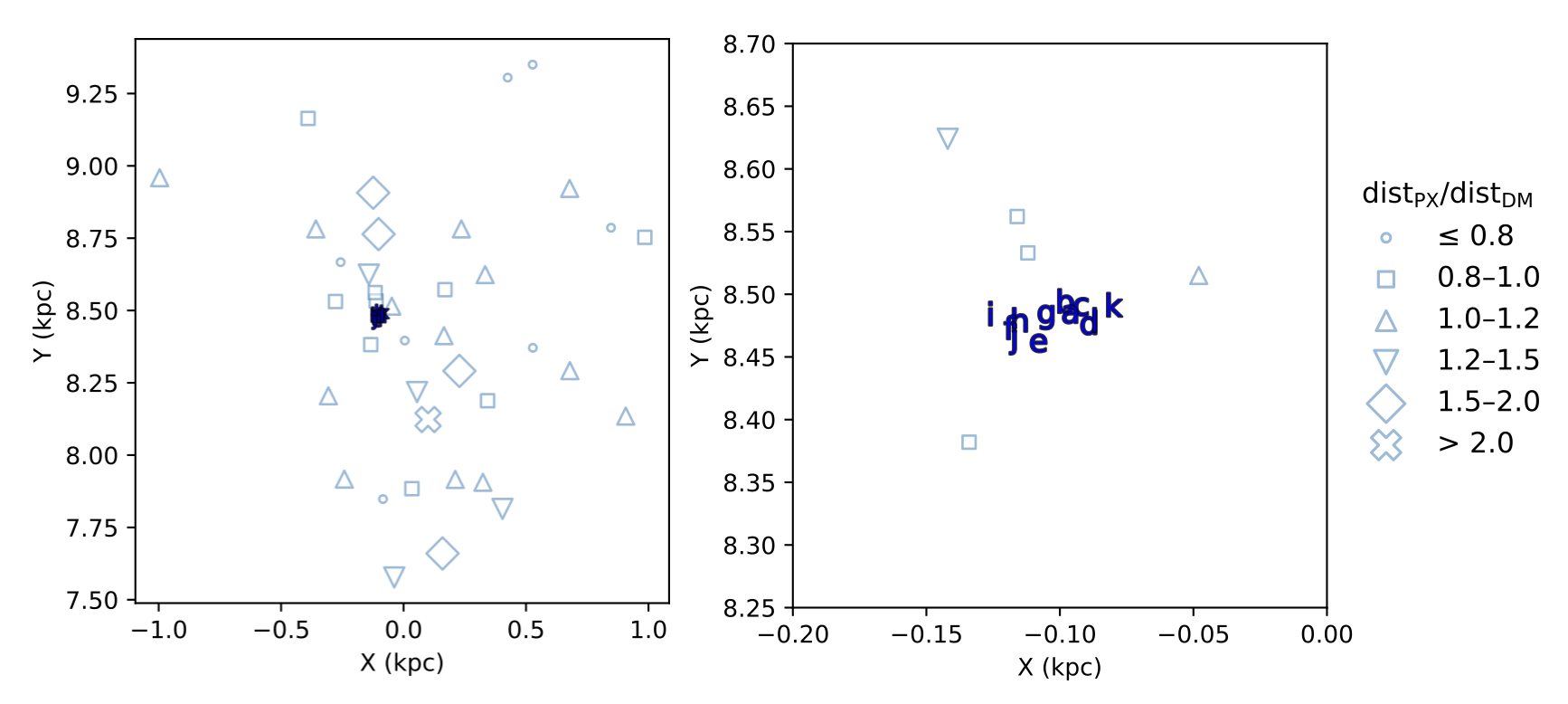}
    \includegraphics[width=0.99\linewidth]{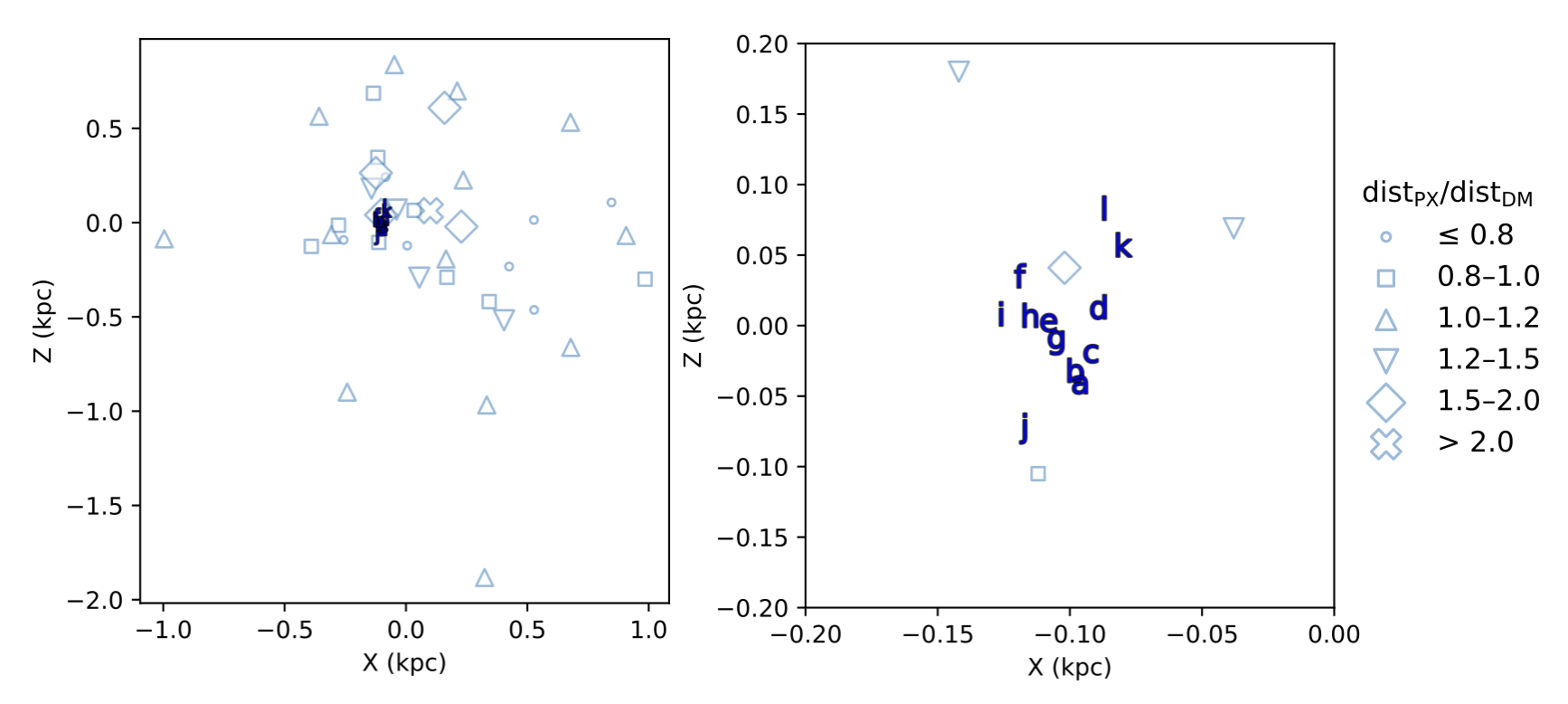}
    \includegraphics[width=0.99\linewidth]{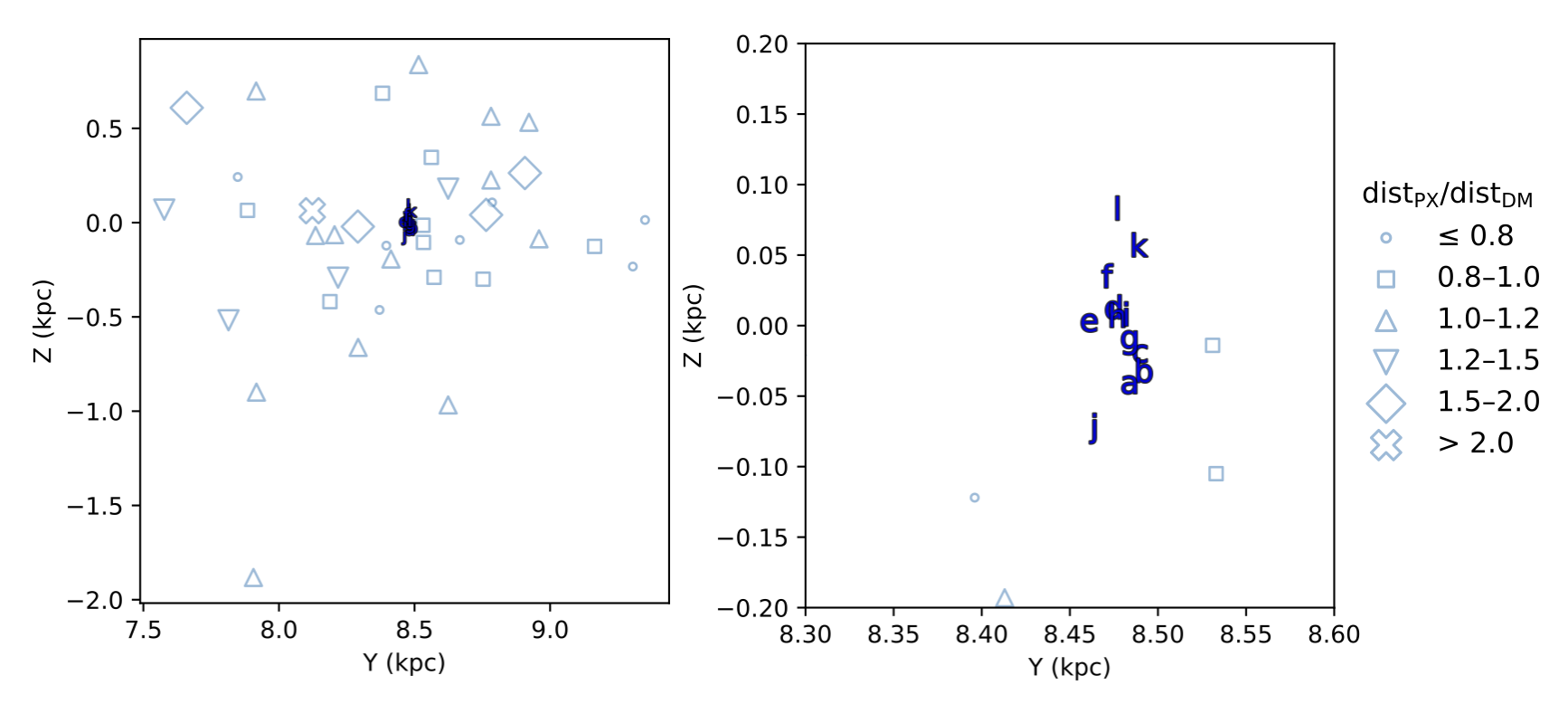}
    \caption{Projected pulsar positions in Galactocentric rectangular coordinates. The left column shows the side–view (X–Y), top–view (X–Z) and front–view (Y–Z) slices. Pulsars with reported parallax distances are classified according to the ratio of their distance estimates from parallax and from the YMW16 model(shown by white symbols), $\mathrm{dist_{PX}}/\mathrm{dist_{DM}}$, and nearby candidates without measured parallaxes are marked by red labels “a”–“l.” 
   The right‐hand panels are zoomed in on the same regions to highlight those labeled sources.
   }
    \label{fig:Map in galactocentric coords}
\end{figure}

Next, we re-display pulsar positions from Fig.~\ref{fig:1.1kpc} in galactocentric right-handed rectangular coordinates in Fig.~\ref{fig:Map in galactocentric coords}.
Here the X axis is directed toward the Galactic Center, the Y axis spans longitude and the Z axis spans latitude.


\newpage
\bibliographystyle{JHEP.bst}
\bibliography{closestns}
\end{document}